\documentclass{article}
\usepackage{spconf,amsmath,graphicx}
\usepackage{diagbox}
\usepackage{stfloats}
\usepackage{booktabs}
\usepackage{bm}
\usepackage{amssymb}
\usepackage{tikz}
\usepackage[colorlinks,linkcolor=red]{hyperref}
\usepackage{cleveref}

\newcommand\copyrighttext{
  \footnotesize \textcopyright 2023 IEEE.  Personal use of this material is permitted.  Permission from IEEE must be obtained for all other uses, in any current or future media, including reprinting/republishing this material for advertising or promotional purposes, creating new collective works, for resale or redistribution to servers or lists, or reuse of any copyrighted component of this work in other works.
}
  
\newcommand\mycopyrightnotice{
\begin{tikzpicture}[remember picture,overlay]
\node[anchor=south,yshift=10pt] at (current page.south) {\fbox{\parbox{\dimexpr\textwidth-\fboxsep-\fboxrule\relax}{\copyrighttext}}};
\end{tikzpicture}
}

\title{Diverse and Vivid Sound Generation from Text Descriptions}
\name{Guangwei Li, Xuenan Xu, Lingfeng Dai, Mengyue Wu$^{\dag}$, Kai Yu$^{\dag}$\thanks{$\dag$ Mengyue Wu and Kai Yu are the corresponding authors.This work has been supported by Shanghai Municipal Science and Technology Major Project (2021SHZDZX0102) and Jiangsu Technology Project (No.BE2022059-2).}}
\address{
X-LANCE Lab, Department of Computer Science and Engineering\\
MoE Key Lab of Artificial Intelligence\\
AI Institute, Shanghai Jiao Tong University, Shanghai, China}
\begin{document}
\ninept
\maketitle
\begin{abstract}
Previous audio generation mainly focuses on specified sound classes such as speech or music, whose form and content are greatly restricted.
In this paper, we go beyond specific audio generation by using natural language description as a clue to generate broad sounds.
Unlike visual information, a text description is concise by its nature but has rich hidden meanings beneath, which poses a higher possibility and complexity on the audio to be generated.
A Variation-Quantized GAN is used to train a codebook learning discrete representations of spectrograms.
For a given text description, its pre-trained embedding is fed to a Transformer to sample codebook indices to decode a spectrogram to be further transformed into waveform by a melgan vocoder.
The generated waveform has high quality and fidelity while excellently corresponding to the given text.
Experiments show that our proposed method is capable of generating natural, vivid audios, achieving superb quantitative and qualitative results.
\end{abstract}

\mycopyrightnotice
\section{Introduction}
\label{sec:intro}

\begin{figure}[!tbp]
  \centering
  \includegraphics[width=0.8\linewidth]{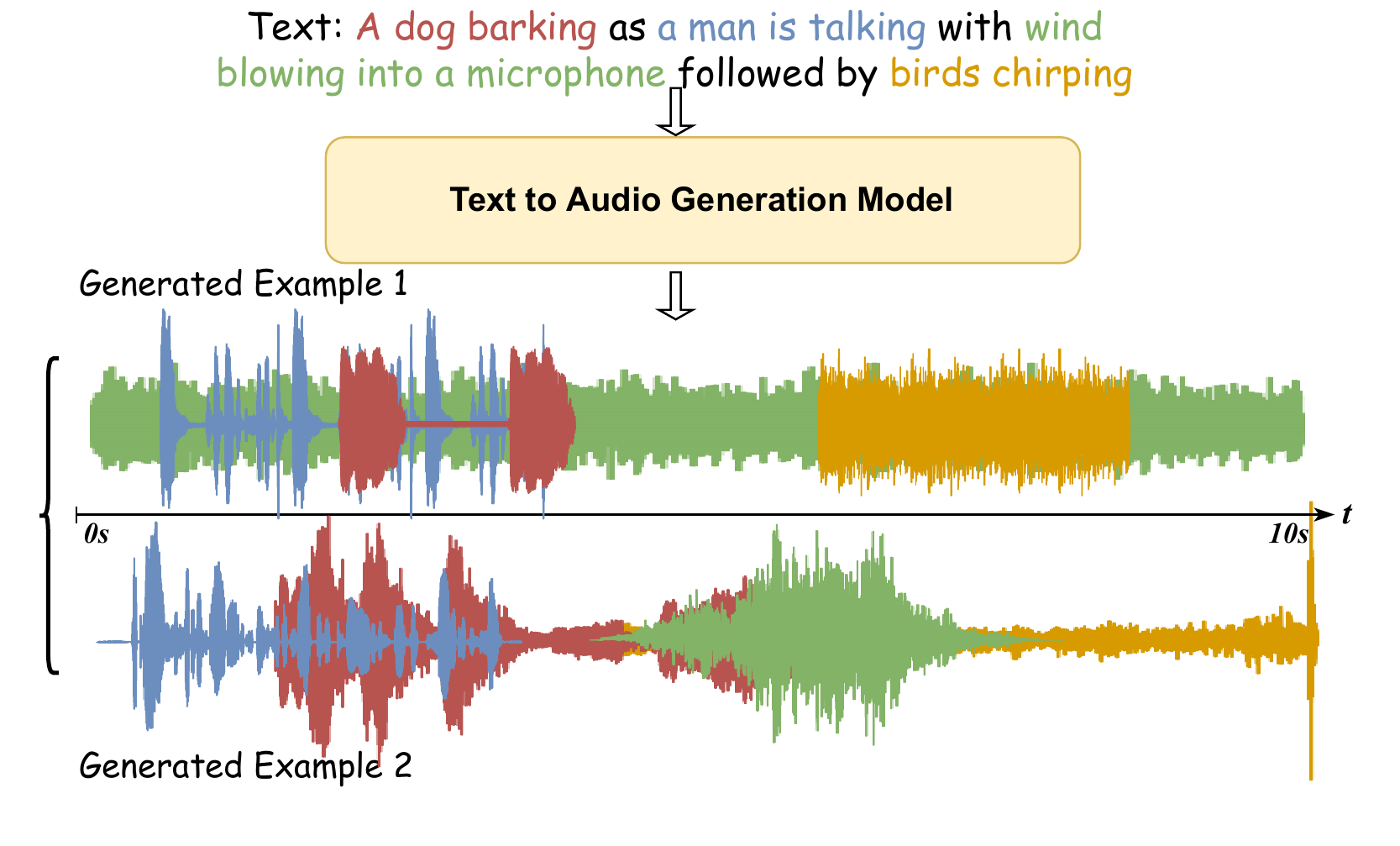}
  \caption{The proposed audio generation from in-the-wild text aims at generating high-quality rich audio content based on any given text. Here are two schematic examples generated from one single sentence which contains multiple sound events from our model.}
  \label{fig:teaser}
\end{figure}

Cross-modality generation task has received increasing amount of attention, spanning from text-image ~\cite{esser2021taming,nichol2021glide,saharia2022photorealistic}, image-audio ~\cite{iashin2021taming,su2020audeo,tan2020spectrogram}, and vice versa. 
However, audio generation from text, in particular from sentence-level descriptions, is still at an emerging stage. 
Audio generation has exhibited unique challenges due to sound events' time-frequency correlations in the real world.
For example, for the sentence `Food is frying, and a woman talks,' the uncertainties in the chronological order and duration of both sounds largely influence the possible audio content. 
In this paper, we define a task in which \textit{\textbf{complex audios}} are generated based on \textit{\textbf{free textual descriptions}} (see \Cref{fig:teaser}). 
The generated audios are expected to maintain high quality and naturalness as well as comprehensively reflect the sound classes and their relations described in the sentences.

\begin{figure*}[!ht]
  \centering
  \includegraphics[width=0.8\linewidth]{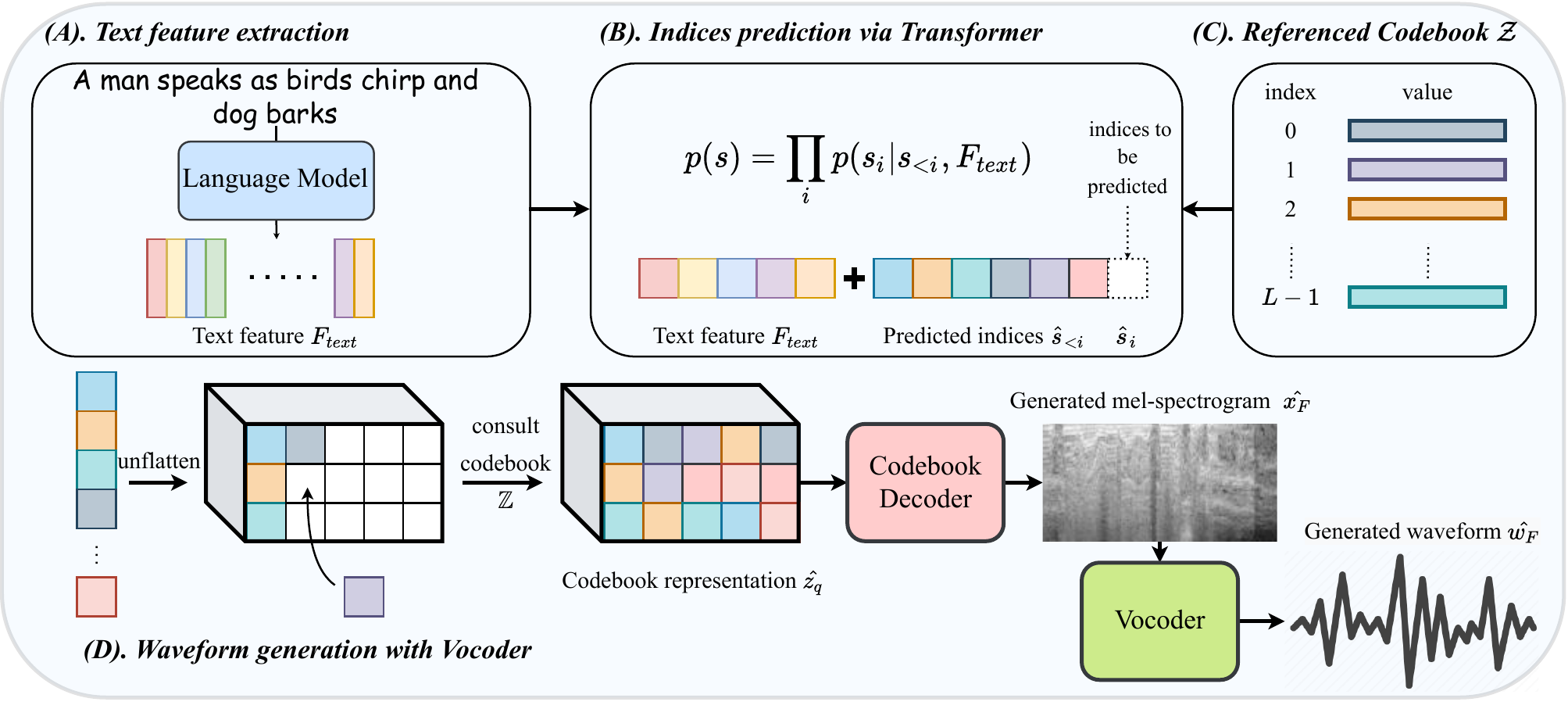}
  \caption{The main structure of our text-to-audio generation system. There are four major procedures when sampling. The text feature is extracted from the given sentence through a pre-trained language model in (A). The codebook indices are generated autoregressively given the text feature in (B). (C) uses a pre-trained codebook to fetch the codebook representation with the indices generated in (B). The representation is subsequently transformed into a mel-spectrogram, which is turned into a waveform through a vocoder in (D).}
  \label{fig:Approach}
  \vspace{-0.4cm}
\end{figure*}

Previous audio generation tasks mainly focus on synthesizing specified sound classes like speech~\cite{shen2018natural,li2018close,ren2020fastspeech} and music~\cite{yang2017midinet,dong2018musegan,wang2019performancenet}.
However, these tasks can be further extended: i) The sound class to generate is pre-defined hence the input is restricted; ii) The content generated is, in fact, a translation of the input modality. 
An emerging study~\cite{okamoto2021onoma} focuses on audio generation given text as input, using onomatopoeic words to synthesize environmental sounds. However, the diversity is still restricted since onomatopoeic words are limited and sometimes less expressive.
Recently, vivid audio generation with less restricted input is implemented with visual cues~\cite{iashin2021taming,su2020audeo,tan2020spectrogram} or onomatopoeic words~\cite{okamoto2021onoma}. 
Yet, there is little research on how to generate diverse audios based on free-text descriptions. Concurrent work focusing on this task includes ~\cite{yang2022diffsound,kreuk2022audiogen}.
Unlike audio-video or audio-image pairs that can be naturally obtained from the original video sources, audio-text pairs need additional annotations.
Annotations describing the same audio may greatly vary from each other in word choices, overall style and writing order. 

Inspired by studies on audio generation with visual cues, we propose a text-to-audio generation system that can generate high-quality audios given descriptive captions.
The system is presented in \Cref{fig:Approach}.
A Transformer and a reference codebook generates dense representations according to the text feature extracted from a pre-trained language model.
Then a decoder transforms the codebook representations into a spectrogram.
Finally the spectrogram is transformed into a waveform with a MelGAN~\cite{kumar2019melgan} vocoder.
The codebook and the decoder are from a pre-trained Variation-Quantized Generative Adversarial Network (VQGAN)~\cite{iashin2021taming,esser2021taming}.
The system is trained on audio captioning datasets~\cite{drossos2020clotho,kim2019audiocaps}, where audio clips and their corresponding text descriptions are provided.


We apply different metrics, including PSNR, FID, MMKL and SPICE, CIDEr (see details in \Cref{sec:dataset_metrics}), as well as subjective human analysis for evaluation.
Both quantitative and qualitative results demonstrate that the generated audios retain high hearing quality and are reflective of the text given, thus, exhibiting the effectiveness of our text-to-audio generation approach. We also present a few generated examples for a more intuitive presentation\footnote{\url{https://ligw1998.github.io/audiogeneration.html}}. 

\section{Text-to-Audio Generation}
\label{sec:method}

Our main objective is to generate audios that can comprehensively reflect the audible elements in the text description given and maintain high quality and fidelity in the meantime.
The following sections introduce the main parts accordingly, with a framework overview illustrated in Figure \ref{fig:Approach}.

\subsection{Discrete spectrogram codebook of learned representations}
\label{sec:discrete}
As mentioned above, our goal is first to construct the audio mel-spectrogram based on the input text.
One straightforward way is to predict the optimal value for every single pixel in the spectrogram.
However, the scale of pixels from the raw spectrogram may be tremendous, making it non-operative for a Transformer to attend.

In order to express the mel-spectrogram more efficiently, we use a discrete codebook to represent the spectrogram.
Our approach refers to a VQGAN to train the codebook, which was initially proposed in ~\cite{esser2021taming} and adapted to audio in ~\cite{iashin2021taming}.
VQGAN is an autoencoder that has a smaller representation size than vector-quantized variational autoencoder (VQVAE)~\cite{van2017neural} and is originally meant to decode an image. 
Here we adopt VQGAN to reconstruct a given spectrogram \( x \in R^{M\times{T}}\) to \( \hat{x}\in R^{M\times{T}}\), \(M\) serves as the number of mel banks while \(T\) is the time dimension.

During training, the spectrogram is encoded into a smaller \( \hat{z} \in \mathbb{R}^{m\times{t}\times{n_z}}\) with an encoder \( E\), where \(n_z\) serves as the dimensional of codes.
The small-scale representation can then be flattened into an \( m \times{t} \) sequence to determine the entries to be chosen from the learned codebook \(\mathcal{Z}=\{z_l\}^{L}_{l=1}\subset \mathbb{R}^{n_z}\).
We use an element-wise quantization \(q\) to map each individual \(\hat{z}_{ij}\) in \(\hat{z}\) into its closest codebook entry \(z_l\), which can be described as:
\begin{equation}
\label{eq:quantize}
    z_q = q(\hat{z}) := \mathop{\arg\min}\limits_{z_{l}\in \mathcal{Z}}{\left\Vert \hat{z}_{ij}-z_l \right\Vert}
\end{equation}
\(z_q\) is the quantized representation and can be further decoded into a spectrogram \(\hat{x}\) by a decoder \(D\):
\begin{equation}
\label{eq:reconstruct}
    \hat{x}=D(z_q)=D(q(\hat{z}))
\end{equation}
Our goal is to minimize the difference between \(x\) and \(\hat{x}\).

Although the quantization process in \Cref{eq:quantize} is nondifferentiable, according to ~\cite{bengio2013estimating}, the gradient for back-propagation in \Cref{eq:reconstruct} can be copied to the decoder from the encoder by a gradient estimator.
The loss function of \Cref{eq:reconstruct} can be described as:
\begin{equation}
\label{eq:lossvq}
    \mathcal{L}_{VQ}(E,D,\mathcal{Z})={\left\Vert \hat{x}-x \right\Vert}^2+{\left\Vert \rm{sg}[\hat{z}]-z_q \right\Vert}_2^2 +\beta {\left\Vert \rm{sg}[z_q]-\hat{z} \right\Vert}_2^2
\end{equation}
where \(\mathcal{L}_{rec}={\left\Vert \hat{x}-x \right\Vert}^2\) and \(\rm{sg} [\cdot] \) stands for the stop gradient operation that has zero gradient during back-propagation and \(\beta\) is a weighting factor.

VQGAN extents the loss function in \Cref{eq:lossvq} with discriminator~\cite{isola2017image} and perceptual loss~\cite{johnson2016perceptual}. We follow ~\cite{iashin2021taming} and adapt the VQGAN on the spectrogram, and replace the original perceptual loss with a learned perceptual audio patch similarity. The final loss function of Spectrogram VQGAN is:
\begin{equation}
\label{eq:loss}
\begin{aligned}
    \mathcal{L}_{SpecVQGAN}(\{E,D,\mathcal{Z}\},C)=\mathcal{L}_{VQ}(E,D,\mathcal{Z})+ logC(x)\\
    +log(1-C(\hat{x}))+\sum_{s}{\frac{1}{M^s T^s}}{\left\Vert x^s-\hat{x}^s \right\Vert}_2^2 
\end{aligned}
\end{equation}
Here, \(C\) is a patch-based discriminator and \(x^s\) and \(\hat{x}^s\) are features of ground truth and reconstructed spectrogram from the \(s^{th}\) scale of VGGish-ish~\cite{iashin2021taming} trained on AudioSet.


\subsection{Text feature extraction}
\label{sec:textfeat}
Before training a Transformer to sample the codebook indices, the text description needs to be extracted to provide the necessary information to generate an audio clip.
The sentence is fed to a \textbf{BERT}-based~\cite{devlin2018bert} model to be transformed into an embedding sequence.
We also use the text encoder from \textbf{CLIP}~\cite{radford2021learning}.
Inspired by CLIP, we also perform audio-text pre-training with the contrastive learning paradigm.
After pre-training, the text encoder is taken as an advanced feature extractor, which is called \textbf{CL`A'P}~\cite{xu2022sjtu}. Similar to CLIP, within each batch, we take the matching audio-text pairs as positive samples, while the mismatched pairs as negative samples. We train the model on an integrated dataset based on currently available public text-audio dataset (AudioCaps, Clotho, MACS) .

\subsection{Text conditioned generation with Transformer}
\label{sec:transformer}

With the encoder \(E\) and decoder \(D\) trained in Section Discrete and the text feature extracted in \Cref{sec:textfeat}, we are able to generate a spectrogram with a latent Transformer.
Given the text feature \(F_{text}\), the Transformer learns to predict the distribution of the next indice \(s_i\) with the indices \(s_{<i}\).
The likelihood of the sequence \(p(s)\) is:

\begin{equation}
\label{eq:likelihood}
    p(s|F_{text})=\prod_{i} p(s_{i}|s_{<i},F_{text})
\end{equation}

The Transformer is trained by maximizing the log-likelihood of the conditioned sequence:

\begin{equation}
\label{eq:losstransformer}
    \mathcal{L}_{Transformer}=\mathbb{E}_{x\sim{p(x)}}[-logp(s|F_{text})]
\end{equation}


During inference, the codebook indices \(\hat{s_i}\) are sampled from the distribution provided by \(p\).
Generated autoregressively until \(i=m\times{t}\), the indices are unflattened into shape \((m,t)\) in a column-major way according to ~\cite{iashin2021taming} and used to look up the codebook by replacing each code by its index in codebook \(\mathcal{Z}\):

\begin{equation}
\label{eq:lookup}
    (z_{q})_{ij} = z_l\ \rm{if}\ s_{ij}=l
\end{equation}

where \(i\in{\{1,2,3,...,m\}}\) and \(j\in{\{1,2,3,...,t\}}\).
By mapping the indices back to codebook representation \(\hat{z_q}\in \mathbb{R}^{m\times{t}\times{n_z}}\), the generated spectrogram can be recovered with decoder \(D\):

\begin{equation}
\label{eq:decode}
    \hat{x_{F}}=D(\hat{z_{q}})
\end{equation}

\begin{figure*}[!htbp]
  \centering
  \includegraphics[width=0.8\linewidth]{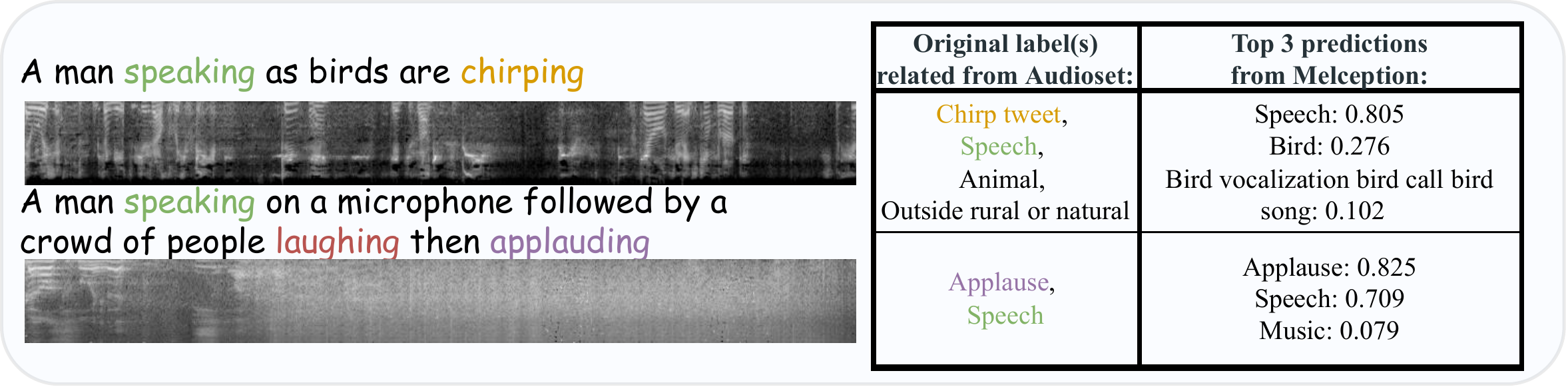}
  \caption{Examples from our text-to-audios generation system. We feed the generated spectrogram into a Melception audio classifier, the event detection results are listed on the right, along with the referenced ground truth label(s) from original audios in AudioSet.}
  \label{fig:audios}
  \vspace{-0.4cm}
\end{figure*}




\section{Experiments}
\label{sec:experiments}

\subsection{Datasets and evaluation metrics}
\label{sec:dataset_metrics}
We conduct our experiments on two datasets: AudioCaps~\cite{kim2019audiocaps} and Clotho~\cite{drossos2020clotho}, and evaluate
the performance objectively and subjectively.

\textbf{Objective Evaluation}
We incorporate several metrics to evaluate the generation performance.
FID~\cite{heusel2017gans}, PSNR and MMKL are used to measure the generated spectrogram quality.
Besides, a high-quality audio generated from text should be able to generate back to captions through an audio captioning model~\cite{xu2022sjtu}. 
We calculate the similarity between the input text and the back-generated one in terms of SPICE~\cite{anderson2016spice} and CIDEr~\cite{vedantam2015cider} to measure the relevance between the input text and the generated audio.



Multiclass Melception-based KL-divergence (MMKL), an improved version of MKL, is initially proposed in ~\cite{iashin2021taming}.
The original classifier Melception is trained on VGGSound with only one sound label per clip, while most of the sound clips in AudioCaps have more than two sound classes, we replace the Softmax activation layer with Sigmoid and train the Melception using the binary cross entropy loss on AudioSet for evaluation. 


\textbf{Subjective Evaluation}
We adopt human evaluation on both the \textit{quality} and \textit{relevance} of the generated waveforms.
For each text description given, the rater will score on both the relevance and the quality of the audio generated, with a 0-10 scale by four raters.
More specifically, `relevance' measures how the audio matches the text description and `quality' indicates how the sound(s) generated is similar to sound(s) in real life and how difficult it is to identify them.

\subsection{Experimental setup}
\label{sec:setup}

For the data pre-processing, each audio clip is \(\approx\)10s long of 32000Hz, and is extracted every 12.5ms with a window size of 50ms.
The number of mel banks is 80.
We utilize different pre-trained language models, including BERT-base-uncased, BERT-medium, CLIP and CL`A'P text encoders, to transform the text description into a feature taken from hidden states.
For all text encoders, the maximum sequence length is set to 30.

When training a codebook, we set the length of the codebook \(|\mathcal{Z}|\) as 1024. 
The adversarial part of the loss in \Cref{eq:loss} is zeroed-out in the initial stage of the codebook training process to stabilize the model.
For the Transformer part, we use a GPT2-based architecture~\cite{lagler2013gpt2}, with a hidden dimension of 1024. 
The output of the Transformer goes through a 1024-way classifier ending with Softmax, which learns a distribution over the next codebook indice.
During inference, we clip the distribution of the next codebook token to Top-K probabilities.
It provides a control for the sample’s diversity.
K ranges from K = \(|\mathcal{Z}|=1024\) to K = 1, here we choose K=512.
The Melception for MMKL is trained on AudioSet and achieves an mAP of 33.4\%.
A MelGAN~\cite{kumar2019melgan} based architecture is used to bridge the gap between the predicted mel-spectrogram and waveform.


\section{Results}
\label{sec:results}
In this section, we evaluate the generated audio samples according to the objective and subjective metrics on AudioCaps and Clotho. 


\begin{table*}[!htbp]
  \caption{The performance of systems under different configurations.}
  \label{tab:results}
  \centering
   {%
    \begin{tabular}{ l| c||c | c |c |c | c}
      \toprule
      Condition& Text Encoder &PSNR$\uparrow$ & FID$\downarrow$ & MMKL$\downarrow$ & SPICE$\uparrow$ & CIDEr$\uparrow$\\
      \midrule
      (a) No Feat      &\diagbox{}{}                &   13.69  & 0.977 &  15.80  & 0.035 & 0.053\\
      \midrule
      (b) 1 Feat       & BERT-medium             &   14.62  & 0.931 &  10.98 & 0.073& 0.180\\
      (c) 1 Feat       & BERT-base             &   14.78  & 0.936 &  10.43  &0.075 &0.183\\
      (d) 1 Feat       & CLIP             &   14.89  & 0.918 &  10.00  & 0.083 & 0.194\\
      (e) 1 Feat       & CL`A'P             &  \textbf{14.97}    & 0.784 &  8.74  & 0.092& 0.246\\
      \midrule
      (f) 30 Feats      & BERT-medium          &   14.66  & 0.858 & 10.06  &0.081 &0.190\\
      (g) 30 Feats      & BERT-base                &   14.92  & 0.851 & 10.05 & 0.085& 0.210\\
      (h) 30 Feats      & CL`A'P                 &   14.83  & \textbf{0.778} & \textbf{8.57}  & \textbf{0.096} &\textbf{0.249}\\
      \bottomrule
    \end{tabular}%
   }
\vspace{-0.4cm}
\end{table*}

\subsection{Quantitative results}
\label{sec:quantitative}
\textbf{AudioCaps} Our final results are calculated to evaluate the similarity between the generated and the ground-truth spectrograms, illustrated in \Cref{tab:results}.
Since there is no previous work on text-to-natural audio generation, we compare our best model (h) with models using different text encoders mentioned in \Cref{sec:textfeat}.
We also alter the length of the feature sequence fed to the model as previous work suggests the effectiveness and practicability of reduced features~\cite{devlin2018bert}. 
Current feature length comparison includes \texttt{No Feat} (0 feature by replacing the text feature with a random embedding), \texttt{1 Feat} (1 feature from the [CLS] token, output embedding from the first hidden state) and \texttt{30 Feats} (full output features).
(a) listed our results without any input condition.
We set this as an upper bound for MMKL; (b, c, d, e) compares on 1 Feat condition, the influence of different text feature extractors; similarly, (f, g, h) concentrates on the entire 30 feats with different feature sizes.

Several observations can be extracted from the results: 1) if sampling without input text conditions, the upper bound of PSNR and MMKL is 13.69 and 15.80, respectively, with the captioning metrics rather low, in (a); 2) Adding at least one text feature will lead to a significant shift in these two metrics to 14.78$\uparrow$ and 10.43$\downarrow$ (c) and a boost in captioning results, suggesting the importance of the input text information to the generated audios; 
3) An increase in the number of text features will increase the relevance (lower MMKL) as well as the overall generation quality (lower FID) (f,g,h).
We hence infer that with more input information, the model is better capable of differentiating the sound sources in the input sentence and the relationship.
4) Using text features with smaller feature sizes extracted from a smaller language model will slightly affect the generation quality negatively (d,e)(b,c). 
However, this is not a critical factor since generating audio content from a sentence focuses most on the sound classes mentioned and their temporal relations of occurrence.
5) Our Best model is (h) using CL`A'P, which means with audio-text contrastive learning, the features extracted are better at representing audio events from the original text given.
Visualized sample spectrograms along with their audio tagging results from Melception are illustrated in \Cref{fig:audios}.

\textbf{Clotho} 
Generally, the overall audio generation performance on Clotho, as shown in \Cref{tab:clotho results}, is inferior to the performance on AudioCaps, since the dataset size of Clotho is distinctively smaller than AudioCaps and the text descriptions in Clotho are more literary.
More complex syntax and terms are utilized in Clotho than AudioCaps, making it more difficult for the model to parse and build connections between text and audio.
If we use the first-stage codebook trained on AudioCaps and train the Transformer on Clotho, the result turns better, suggesting a stronger and more effective codebook is trained on AudioCaps.

\begin{table}[!htbp]
  \caption{Results on Clotho test set, with codebooks trained on two datasets.}
  \label{tab:clotho results}
  \centering
  \begin{resizebox}{0.7\columnwidth}{!}
   {%
    \begin{tabular}{ l||c | c |c }
      \toprule
      Codebook&  MMKL$\downarrow$ & SPICE$\uparrow$ & CIDEr$\uparrow$\\
      \midrule
      Clotho      & 11.613 & 0.049     & 0.091 \\
      AudioCaps       & 10.988 & 0.068 & 0.137 \\
      \bottomrule
    \end{tabular}%
   }
\end{resizebox}
\vspace{-0.2cm}
\end{table}

\subsection{Qualitative analysis}
\label{sec:qualitative}

We randomly picked 200 samples from the 964 generated ones from the AudioCaps test set. 
For every text description, four audios are held out to the raters:
the original audio from AudioCaps; the generated audio with no text feature given; audio conditioned on one feat and on full feats. 
The averaged results are listed in \Cref{tab:human}.

\begin{table}[hbp]
  \caption{Human evaluation results. The generated audio clips are scored on \textit{Relevance} and \textit{Quality} with a rating scale of 0 - 10.}
  \label{tab:human}
  \centering
  \begin{resizebox}{0.7\columnwidth}{!}
   {%
    \begin{tabular}{ l||c | c }
      \toprule
      Condition& Avg. Relevance(/10)&Avg. Quality(/10)\\
      \midrule
      Original                       & \textbf{9.55} &  \textbf{9.55}  \\
      No Feat                       & 1.85 & 4.58  \\
      BERT-b                     & 5.50 &    6.08 \\
      CL`A'P                     & \textbf{6.79} & \textbf{6.13}  \\
      \bottomrule
    \end{tabular}%
   }
   \end{resizebox}
   \vspace{-0.2cm}
\end{table}

It can be observed that both the relevance and the quality of the generated audios are correlated with the conditioned text features.
With the audio pre-trained CL`A'P extractor, the auditory feedback of the generated samples excels the one from other extractors, which is consistent with the quantitative results listed in the previous section. 
However, a considerable gap is observed between the generated audio and the original audio clips.
After consulting the raters, we summarized a few limitations of the current audios generated: 1) different sound classes provided in the text description may occur alternately, which does not happen commonly in real life; 2) the volume of different sounds in a single clip may vary greatly;  3) some rare sound classes are sometimes omitted; 4) the content of the generated `speech' is slurred with no exact words.

\section{Conclusion}
\label{sec:conclusion}

In this paper, we focus on a novel and challenging task of generating high-quality yet diverse sound content given a text description.
With a reference codebook trained by VQGAN, our model is able to reconstruct a mel-spectrogram from a more miniature representation efficiently.
A Transformer-based structure is implemented to bridge the gap between the input text feature and the predicted representation when sampling. 
Our best model achieves an average of 14.83, 0.778, and 8.57 in quantitative metrics PSNR, FID, and MMKL, respectively, along with audio caption results (0.096 with SPICE and 0.249 with CIDEr).
Human evaluation also affirms the overall quality and the relevance of the generated audio clips with the given text.
Both results indicate that our proposed approach can generate natural and diversified audios that correspond to the information applied in the text.
We also expect further research, including a better chronological order and duration of the generated sound events.

\bibliographystyle{IEEEbib}
\bibliography{strings,refs}

\begin{thebibliography}{10}

\bibitem{esser2021taming}
Patrick Esser, Robin Rombach, and Bjorn Ommer,
\newblock ``Taming transformers for high-resolution image synthesis,''
\newblock in {\em Proceedings of the IEEE/CVF Conference on Computer Vision and
  Pattern Recognition}, 2021, pp. 12873--12883.

\bibitem{nichol2021glide}
Alex Nichol and et~al. Dhariwal,
\newblock ``Glide: Towards photorealistic image generation and editing with
  text-guided diffusion models,''
\newblock {\em arXiv preprint arXiv:2112.10741}, 2021.

\bibitem{saharia2022photorealistic}
Chitwan Saharia, William Chan, Saurabh Saxena, and et~al. Li,
\newblock ``Photorealistic text-to-image diffusion models with deep language
  understanding,''
\newblock {\em arXiv preprint arXiv:2205.11487}, 2022.

\bibitem{iashin2021taming}
Vladimir Iashin and Esa Rahtu,
\newblock ``Taming visually guided sound generation,''
\newblock {\em arXiv preprint arXiv:2110.08791}, 2021.

\bibitem{su2020audeo}
Kun Su and et~al.,
\newblock ``Audeo: Audio generation for a silent performance video,''
\newblock {\em Advances in Neural Information Processing Systems}, vol. 33, pp.
  3325--3337, 2020.

\bibitem{tan2020spectrogram}
Huadong Tan, Guang Wu, Pengcheng Zhao, and Yanxiang Chen,
\newblock ``Spectrogram analysis via self-attention for realizing cross-model
  visual-audio generation,''
\newblock in {\em ICASSP 2020-2020 IEEE International Conference on Acoustics,
  Speech and Signal Processing (ICASSP)}. IEEE, 2020, pp. 4392--4396.

\bibitem{shen2018natural}
Jonathan Shen, Ruoming Pang, Ron~J Weiss, Rj~Schuster, et~al.,
\newblock ``Natural tts synthesis by conditioning wavenet on mel spectrogram
  predictions,''
\newblock in {\em 2018 IEEE international conference on acoustics, speech and
  signal processing (ICASSP)}. IEEE, 2018, pp. 4779--4783.

\bibitem{li2018close}
Naihan Li, Shujie Liu, Yanqing Liu, Sheng Zhao, Ming Liu, and Ming Zhou,
\newblock ``Close to human quality tts with transformer,''
\newblock {\em arXiv preprint arXiv:1809.08895}, 2018.

\bibitem{ren2020fastspeech}
Yi~Ren, Chenxu Hu, Xu~Tan, Tao Qin, Sheng Zhao, Zhou Zhao, and Tie-Yan Liu,
\newblock ``Fastspeech 2: Fast and high-quality end-to-end text to speech,''
\newblock {\em arXiv preprint arXiv:2006.04558}, 2020.

\bibitem{yang2017midinet}
Li-Chia Yang and et~al. Chou,
\newblock ``Midinet: A convolutional generative adversarial network for
  symbolic-domain music generation,''
\newblock {\em arXiv preprint arXiv:1703.10847}, 2017.

\bibitem{dong2018musegan}
Hao-Wen Dong, Wen-Yi Hsiao, Li-Chia Yang, and Yi-Hsuan Yang,
\newblock ``Musegan: Multi-track sequential generative adversarial networks for
  symbolic music generation and accompaniment,''
\newblock in {\em Proceedings of the AAAI Conference on Artificial
  Intelligence}, 2018, vol.~32.

\bibitem{wang2019performancenet}
Bryan Wang and Yi-Hsuan Yang,
\newblock ``Performancenet: Score-to-audio music generation with multi-band
  convolutional residual network,''
\newblock in {\em Proceedings of the AAAI Conference on Artificial
  Intelligence}, 2019, vol.~33, pp. 1174--1181.

\bibitem{okamoto2021onoma}
Yuki Okamoto and et~al. Imoto,
\newblock ``Onoma-to-wave: Environmental sound synthesis from onomatopoeic
  words,''
\newblock {\em arXiv preprint arXiv:2102.05872}, 2021.

\bibitem{yang2022diffsound}
Dongchao Yang, Jianwei Yu, Helin Wang, Wen Wang, Chao Weng, Yuexian Zou, and
  Dong Yu,
\newblock ``Diffsound: Discrete diffusion model for text-to-sound generation,''
\newblock {\em arXiv preprint arXiv:2207.09983}, 2022.

\bibitem{kreuk2022audiogen}
Felix Kreuk, Gabriel Synnaeve, Adam Polyak, Uriel Singer, Alexandre
  D{\'e}fossez, Jade Copet, Devi Parikh, Yaniv Taigman, and Yossi Adi,
\newblock ``Audiogen: Textually guided audio generation,''
\newblock {\em arXiv preprint arXiv:2209.15352}, 2022.

\bibitem{kumar2019melgan}
Kundan Kumar, Rithesh Kumar, and et~al. de~Boissiere,
\newblock ``Melgan: Generative adversarial networks for conditional waveform
  synthesis,''
\newblock {\em Advances in neural information processing systems}, vol. 32,
  2019.

\bibitem{drossos2020clotho}
Konstantinos Drossos, Samuel Lipping, and Tuomas Virtanen,
\newblock ``Clotho: An audio captioning dataset,''
\newblock in {\em ICASSP 2020-2020 IEEE International Conference on Acoustics,
  Speech and Signal Processing (ICASSP)}. IEEE, 2020, pp. 736--740.

\bibitem{kim2019audiocaps}
Chris~Dongjoo Kim, Byeongchang Kim, Hyunmin Lee, and Gunhee Kim,
\newblock ``Audiocaps: Generating captions for audios in the wild,''
\newblock in {\em Proceedings of the 2019 Conference of the North American
  Chapter of the Association for Computational Linguistics: Human Language
  Technologies, Volume 1 (Long and Short Papers)}, 2019, pp. 119--132.

\bibitem{van2017neural}
Aaron Van Den~Oord, Oriol Vinyals, et~al.,
\newblock ``Neural discrete representation learning,''
\newblock {\em Advances in neural information processing systems}, vol. 30,
  2017.

\bibitem{bengio2013estimating}
Yoshua Bengio and et~al. L{\'e}onard,
\newblock ``Estimating or propagating gradients through stochastic neurons for
  conditional computation,''
\newblock {\em arXiv preprint arXiv:1308.3432}, 2013.

\bibitem{isola2017image}
Phillip Isola, Jun-Yan Zhu, Tinghui Zhou, and Alexei~A Efros,
\newblock ``Image-to-image translation with conditional adversarial networks,''
\newblock in {\em Proceedings of the IEEE conference on computer vision and
  pattern recognition}, 2017, pp. 1125--1134.

\bibitem{johnson2016perceptual}
Justin Johnson, Alexandre Alahi, and Li~Fei-Fei,
\newblock ``Perceptual losses for real-time style transfer and
  super-resolution,''
\newblock in {\em European conference on computer vision}. Springer, 2016, pp.
  694--711.

\bibitem{devlin2018bert}
Jacob Devlin, Ming-Wei Chang, Kenton Lee, and Kristina Toutanova,
\newblock ``Bert: Pre-training of deep bidirectional transformers for language
  understanding,''
\newblock {\em arXiv preprint arXiv:1810.04805}, 2018.

\bibitem{radford2021learning}
Alec Radford, Jong~Wook Kim, Chris Hallacy, Aditya Ramesh, Gabriel Goh,
  Sandhini Agarwal, Girish Sastry, Amanda Askell, Pamela Mishkin, Jack Clark,
  et~al.,
\newblock ``Learning transferable visual models from natural language
  supervision,''
\newblock in {\em International Conference on Machine Learning}. PMLR, 2021,
  pp. 8748--8763.

\bibitem{xu2022sjtu}
Xuenan Xu, Zeyu Xie, Mengyue Wu, and Kai Yu,
\newblock ``The {SJTU} system for {DCASE2022} challenge task 6: Audio
  captioning with audio-text retrieval pre-training,''
\newblock Tech. {R}ep., DCASE2022 Challenge, 2022.

\bibitem{heusel2017gans}
Martin Heusel, Hubert Ramsauer, and et~al. Unterthiner,
\newblock ``Gans trained by a two time-scale update rule converge to a local
  nash equilibrium,''
\newblock {\em Advances in neural information processing systems}, vol. 30,
  2017.

\bibitem{anderson2016spice}
Peter Anderson, Basura Fernando, Mark Johnson, and Stephen Gould,
\newblock ``Spice: Semantic propositional image caption evaluation,''
\newblock in {\em European conference on computer vision}. Springer, 2016, pp.
  382--398.

\bibitem{vedantam2015cider}
Ramakrishna Vedantam, C~Lawrence~Zitnick, and Devi Parikh,
\newblock ``Cider: Consensus-based image description evaluation,''
\newblock in {\em Proceedings of the IEEE conference on computer vision and
  pattern recognition}, 2015, pp. 4566--4575.

\bibitem{lagler2013gpt2}
Klemens Lagler, Michael Schindelegger, Johannes B{\"o}hm, Hana Kr{\'a}sn{\'a},
  and Tobias Nilsson,
\newblock ``Gpt2: Empirical slant delay model for radio space geodetic
  techniques,''
\newblock {\em Geophysical research letters}, vol. 40, no. 6, pp. 1069--1073,
  2013.

\end{thebibliography}

\end{document}